\documentclass[sigconf]{acmart}
\usepackage{wrapfig}
\usepackage{amsmath}
\usepackage{algorithm}
\usepackage{algorithmic}
\usepackage{graphicx}
\usepackage{lipsum}
\usepackage{enumerate}
\usepackage{enumitem}
\usepackage{hyperref}
\usepackage{multirow, makecell}
\usepackage{nccmath}
\usepackage{ragged2e}
\usepackage{diagbox}
\usepackage[caption=false,font=normalsize]{subfig}
\usepackage{threeparttable} 
\usepackage{url}
\usepackage{xspace}
\usepackage{bbding}
\usepackage{comment}
\usepackage{caption}
\usepackage{xcolor}

\AtBeginDocument{%
  \providecommand\BibTeX{{%
    \normalfont B\kern-0.5em{\scshape i\kern-0.25em b}\kern-0.8em\TeX}}}

\setcopyright{acmcopyright}
\copyrightyear{2024}
\acmYear{2024}
\acmDOI{00000.00000}

\acmConference[ACM MobiSys ’24]{In the 22nd ACM International Conference on Mobile Systems, Applications, and Services. }{June 03--07, 2024}{Tokyo, Japan}

\usepackage{etoolbox}
\makeatletter
\patchcmd{\authornote}{\g@addto@macro\addresses{\@authornotemark}}{}{}{}
\makeatother

\newcommand{\eg}{\emph{e.g.},\xspace}

\newcommand{\sysname}{{\sf AdaOper}\xspace}

\ifodd 1

\newcommand\lsc[1]{\textcolor{black}{#1}}

\else

\newcommand\lsc[1]{#1}

\fi

\begin{document}

\title{AdaOper: Energy-efficient and Responsive Concurrent DNN Inference on Mobile Devices}

% \affiliation{\normalsize 
%   Zheng Lin\orcid{0000-0003-4402-1260},$^{1}$ Bin Guo,$^{1}$ Sicong Liu\authornote{Corresponding author}$^{\ast}$,$^{1}$,
%   Wentao Zhou$^2$, Yasan Ding,$^1$ Yu Zhang,$^1$ and Zhiwen Yu$^{1,2}$
% }
% \affiliation{\normalsize 
%   \institution{$^1$Northwestern Polytechnical University, Xi’an, China,$^2$Harbin Engineering University, Harbin, China}
% }

\author{Zheng Lin$^{1}$, Bin Guo$^{1}$, Sicong Liu$^{1}$$^{\ast}$,
   Wentao Zhou$^2$, Yasan Ding$^1$, Yu Zhang$^1$, and Zhiwen Yu$^{1,2}$}

\authornote{Corresponding author: scliu@nwpu.edu.cn}
\affiliation{%
  \institution{$^1$Northwestern Polytechnical University, Xi’an, China,$^2$Harbin Engineering University, Harbin, China}
  \country{}
}

% \author{Zheng Lin}
% \orcid{0000-0002-3644-862X}
% \affiliation{%
%   \institution{Northwestern Polytechnical University}
%   \city{Xi'an}
%   \country{China}
% }

% \author{Bin Guo}
% \orcid{0000-0001-6097-2467}
% \affiliation{%
%   \institution{Northwestern Polytechnical University}
%   \city{Xi'an}
%   \country{China}
% }

% \author{Sicong Liu}
% \authornote{Corresponding author}
% \orcid{0000-0003-4402-1260}
% \affiliation{%
%   \institution{Northwestern Polytechnical University}
%   \city{Xi'an}
%   \country{China}
% }

% \author{Wentao Zhou}
% \orcid{0009-0002-3165-4288}
% \affiliation{%
%   \institution{Harbin Engineering University}
%   \city{Harbin}
%   \country{China}
% }
  
% \author{Yasan Ding}
% \affiliation{%
%   \institution{Northwestern Polytechnical University}
%   \city{Xi'an}
%   \country{China}
% }

% \author{Yu Zhang}
% \affiliation{%
%   \institution{Northwestern Polytechnical University}
%   \city{Xi'an}
%   \country{China}
% }

% \author{Zhiwen Yu}
% \orcid{0000-0002-9905-3238}
% \affiliation{%
%   \institution{Northwestern Polytechnical University}   
%   \city{Xi'an}
%   \country{China}
% }
% \affiliation{%
%   \institution{Harbin Engineering University}
%   \city{Harbin}
%   \country{China}
% }

\begin{abstract}
\lsc{Deep neural network (DNN) has driven extensive applications in mobile technology. 
However, for long-running mobile apps like voice assistants or video applications on smartphones, energy efficiency is critical for battery-powered devices. 
The rise of heterogeneous processors in mobile devices today has introduced new challenges for optimizing energy efficiency. 
Our key insight is that partitioning computations across different processors for parallelism and speedup doesn't necessarily correlate with energy consumption optimization and may even increase it. To address this, we present AdaOper, an energy-efficient concurrent DNN inference system. 
It optimizes energy efficiency on mobile heterogeneous processors while maintaining responsiveness. 
AdaOper includes a runtime energy profiler that dynamically adjusts operator partitioning to optimize energy efficiency based on dynamic device conditions. 
We conduct preliminary experiments, which show that AdaOper reduces energy consumption by 16.88\% compared to the existing concurrent method while ensuring real-time performance.
}
\end{abstract}



\maketitle

\label{sec:intro}
\begin{figure}[t]
  \centering
  \includegraphics[width=0.92\linewidth]{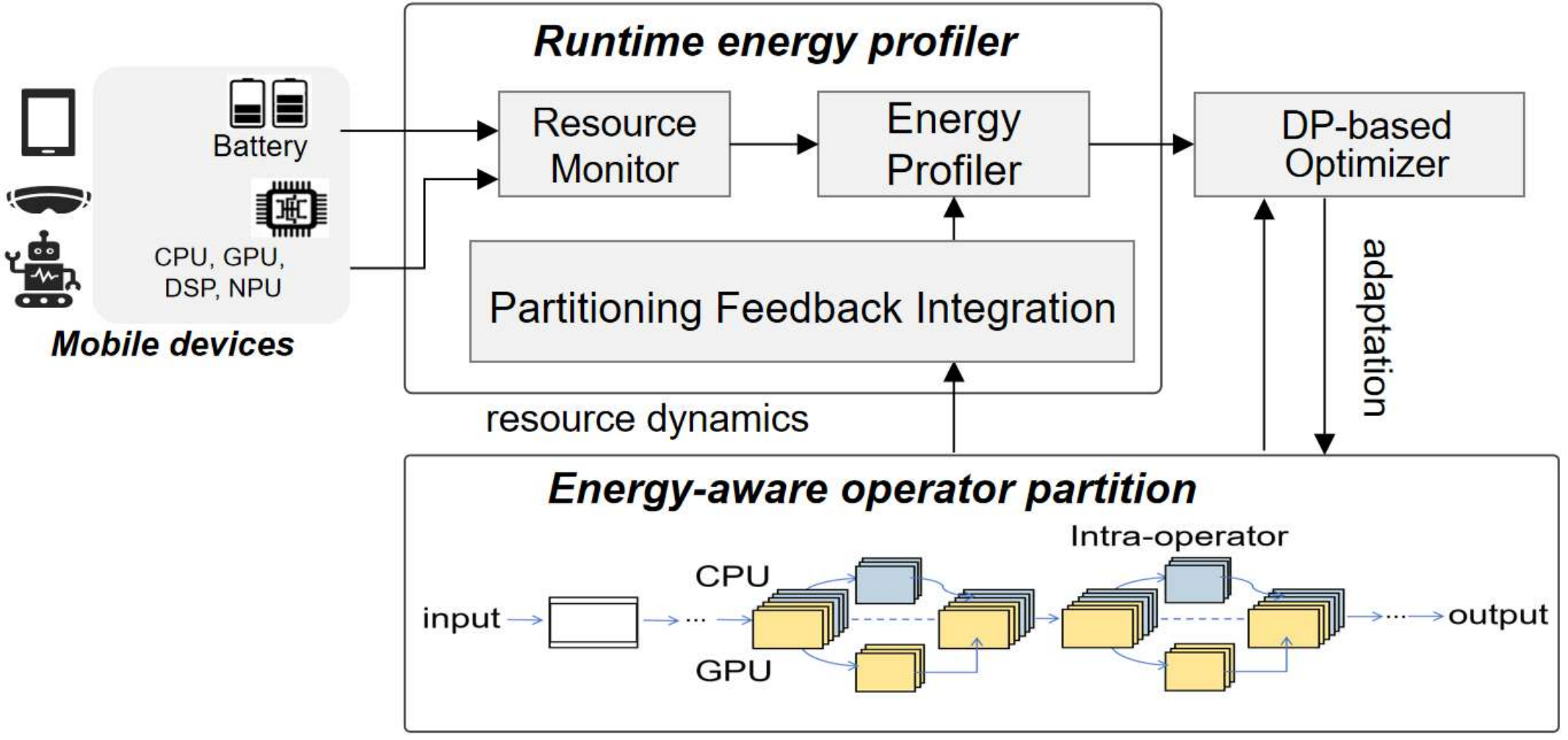}
  \vspace{-2mm}
  \caption{Workflow of AdaOper.}
  \vspace{-6mm}
  \label{fig:overview}
\end{figure}
\section{Introduction}

\lsc{Deep neural networks (DNNs) have fostered numerous successful mobile applications, such as image and speech recognition, autonomous driving, and virtual reality (VR). 
Energy-efficient DNN inference in long-term running mobile applications is necessary, since the majority of mobile devices rely on battery power.
For instance, a smartphone with a 4000mAh battery capacity can typically run for 20 hours under normal usage conditions. However, frequent usage of resource-intensive DL applications such as video detection can impact the user experience by reducing battery duration to approximately 1/3.
To this end, existing research has explored model compression, dynamic voltage and frequency scaling (DVFS), memory access optimization, and arithmetic intensity improvement.}

\lsc{The widespread adoption of heterogeneous processors such as GPU and NPU in mobile devices for accelerating DNN computation brings new challenges for energy optimization. 
Achieving energy-efficient concurrent DNN inference on heterogeneous processors is non-trivial. 
Our \textit{key insight} is that optimizing computation parallelism and latency across heterogeneous processors does not always translate to energy efficiency optimization and may even result in increased energy costs in certain scenarios. 
To tackle this challenge, we present to seek the optimal tradeoff between energy efficiency and speedup in concurrent DNN inference, considering factors like mobile workload characteristics, resource availability, and device constraints. 
However, we face two challenges:
}

\begin{itemize}
    \item \lsc{\textit{Challenge \#1}: Predicting hardware-related energy consumption before exact inference execution is intractable. 
    Existing studies rely on \textit{offline} energy prediction based on regression functions or neural networks\cite{fahad2019comparative}.
    How to incorporate dynamic hardware conditions poses a significant challenge.
    Furthermore, extending \textit{single-processor} energy prediction methods to \textit{cross-heterogeneous-processor} is challenging due to complex factors like cross-processor data communication overhead and the dynamic resources of heterogeneous processors.
    }
    \item \textit{Challenge \#2}: \lsc{Given the runtime energy feedback, realizing \textit{fast} adaptation for optimal DNN computation partitioning at the suitable granularity, \eg operators, is non-trivial. 
    Heterogeneous processors are vulnerable to dynamic system workloads, making it difficult to quickly determine the optimal partitioning strategy from a large search space at runtime. 
    While existing research relies on pre-defined static partitioning or rule-based heuristics, it falls short in addressing the dynamic and real-time adaptation needed for varying workloads and processor conditions.
    }
\end{itemize}

To overcome these challenges, we introduce the design of AdaOper, an energy-efficient and responsive concurrent DNN inference system tailored for mobile devices. 
AdaOper comprises two key modules: the \textit{runtime energy profiler} and the \textit{energy-aware operator partition }module.
Specifically, the runtime energy profiler integrates resource dynamics and workload forecasting. 
This module continuously monitors device states and predicts workload dynamics to provide accurate, real-time energy feedback.
The energy-aware operator partition module autonomously adjusts the partitioning of DNN operators in an energy-aware manner. 
It mainly ensures that the system achieves optimal performance per energy unit consumed, responding to the real-time energy measurements provided by the profiler, thereby optimizing the overall system efficiency.

% To overcome these challenges, we propose the design of AdaOper, an energy-efficient and responsive concurrent DNN inference system on mobile devices. 
% \sysname comprises two modules, \ie \textit{runtime energy profiler} module and\textit{ energy-aware operator partition} module.
% Specifically, the \textit{runtime energy profiler} module realizes real-time energy cost formulation, leveraging a novel fusion of device state and workload forecasting. 
% The \textit{energy-aware operator partition} module autonomously adjusts the DNN operator partition, ensuring optimal performance per energy unit consumed.
% The profiler’s real-time energy measurements dynamically inform the operator partition strategy, enabling an adaptive distribution of operators, responding to dynamic heterogeneous resources.

\section{AdaOper DESIGN}

\subsection{Runtime Energy Profiler}
\lsc{It is non-trivial to timely and accurately predict the runtime energy cost of concurrent DL tasks on mobile devices.
Existing research mainly focuses on offline energy cost formulation, assuming abundant resources, by leveraging coefficient functions of pre-detected computational and I/O loads on different chips, or employing neural networks with fitted mappings. 
However, applying these approaches to cross-chip energy prediction, with increased complexity and dynamics, proves challenging.
}
They inadequately address the dynamic nature of cross-processor conditions and the varied energy consumption patterns of DL tasks. 
To overcome this limitation, we propose a runtime energy profiler.

\lsc{Our profiler combines Gradient Boosting Decision Trees (GBDT) for offline energy consumption modeling and Gated Recurrent Unit (GRU) algorithms for runtime adjustments. 
By leveraging GBDT, it comprehensively evaluates operational factors such as processor frequency and memory bandwidth across diverse mobile devices. 
During runtime, it utilizes a resource monitor and hardware sensors to continuously monitor processor resources and device states, integrating feedback from ongoing DL tasks.
The GRU algorithms dynamically refine the energy model, ensuring responsiveness to both device conditions and task demands. 
This adaptive approach effectively accommodates dynamic changes in mobile device environments and improves the precision of energy feedback.
}

\vspace{-3mm}

\subsection{Energy-aware operator partition}
\lsc{Efficiently executing concurrent DL tasks on mobile devices for speedup while conserving energy poses a significant challenge, especially in cross-heterogeneous-processor environments. 
Previous approaches rely on fixed or latency-efficient computation partitioning strategies, which lack adaptability to the dynamic conditions of mobile devices, such as fluctuations in energy availability and processor loads.
Furthermore, some methods prioritize achieving parallelism and speedup on heterogeneous processors, overlooking the impact on energy consumption. 
This oversight results in suboptimal computing/memory/energy utilization, hindering the overall efficiency of DL execution on mobile devices.
}

\lsc{
To address the challenges outlined, we present an energy-aware operator partitioning module. This module incorporates a dynamic programming algorithm, progressively selecting the operator partition strategy. 
By simplifying the problem to utilize only a few previous states in the operator partition, we optimize space complexity by storing only those states. 
Additionally, we enhance efficiency by converting a recursive top-down dynamic programming solution to an iterative bottom-up scheme. 
This conversion involves refining the redistribution of partial operators triggered by fluctuations in energy consumption, rather than the entire model.
}
\section{PRELIMINARY RESULTS}
\label{sec:intro}
\vspace{-5mm}
\begin{figure}[htbp] % 'h' here allows LaTeX to place the figure right after the section heading
  % \centering % Uncomment this if you want to center the subfigures
  \subfloat[Latency]{
    \includegraphics[width=0.22\textwidth]{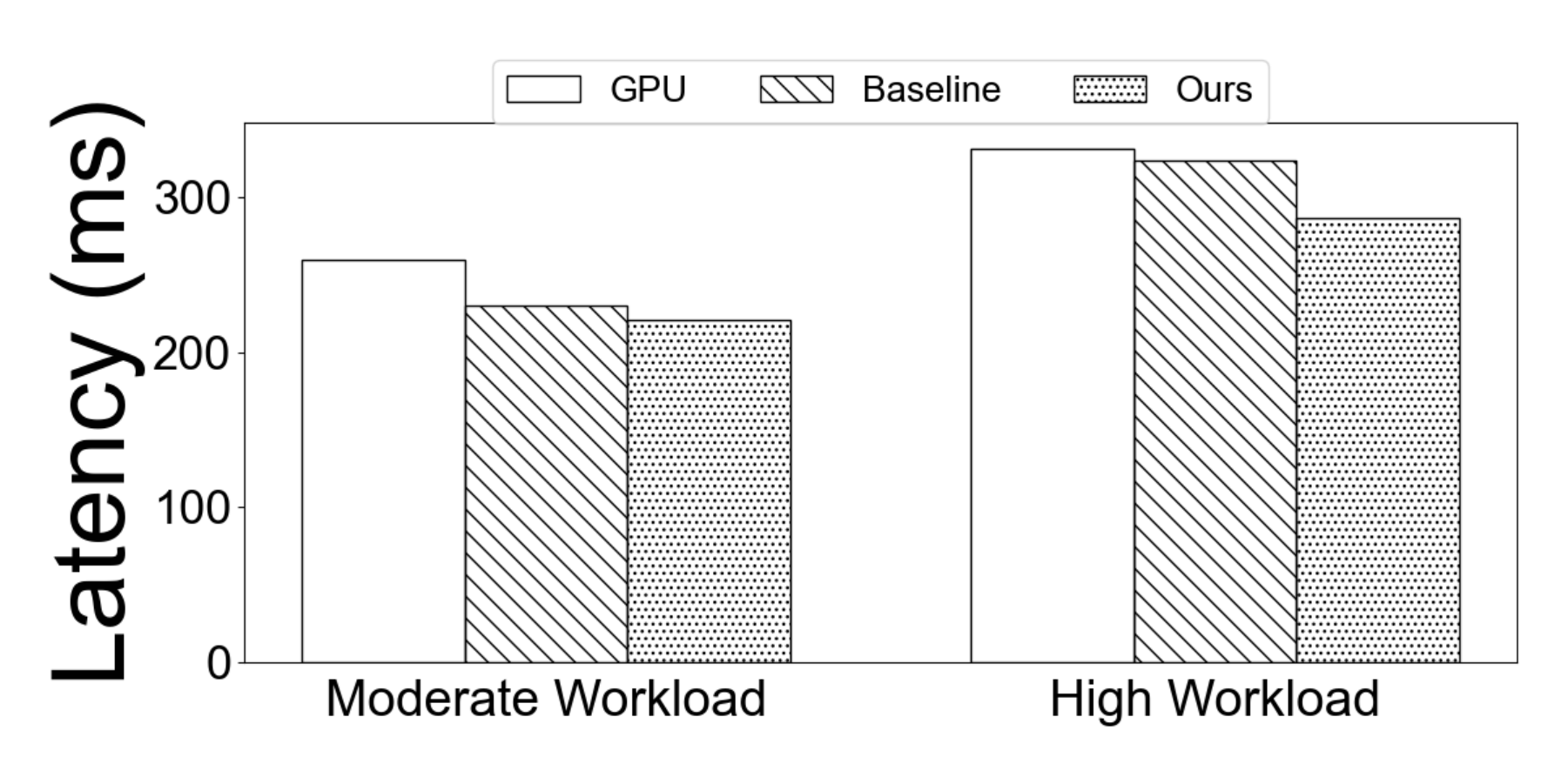}
    \label{fig:Latency and Skew}
  }
  % \hspace{2mm} % Adjust space between the figures if needed
  \subfloat[Energy]{
    \includegraphics[width=0.22\textwidth]{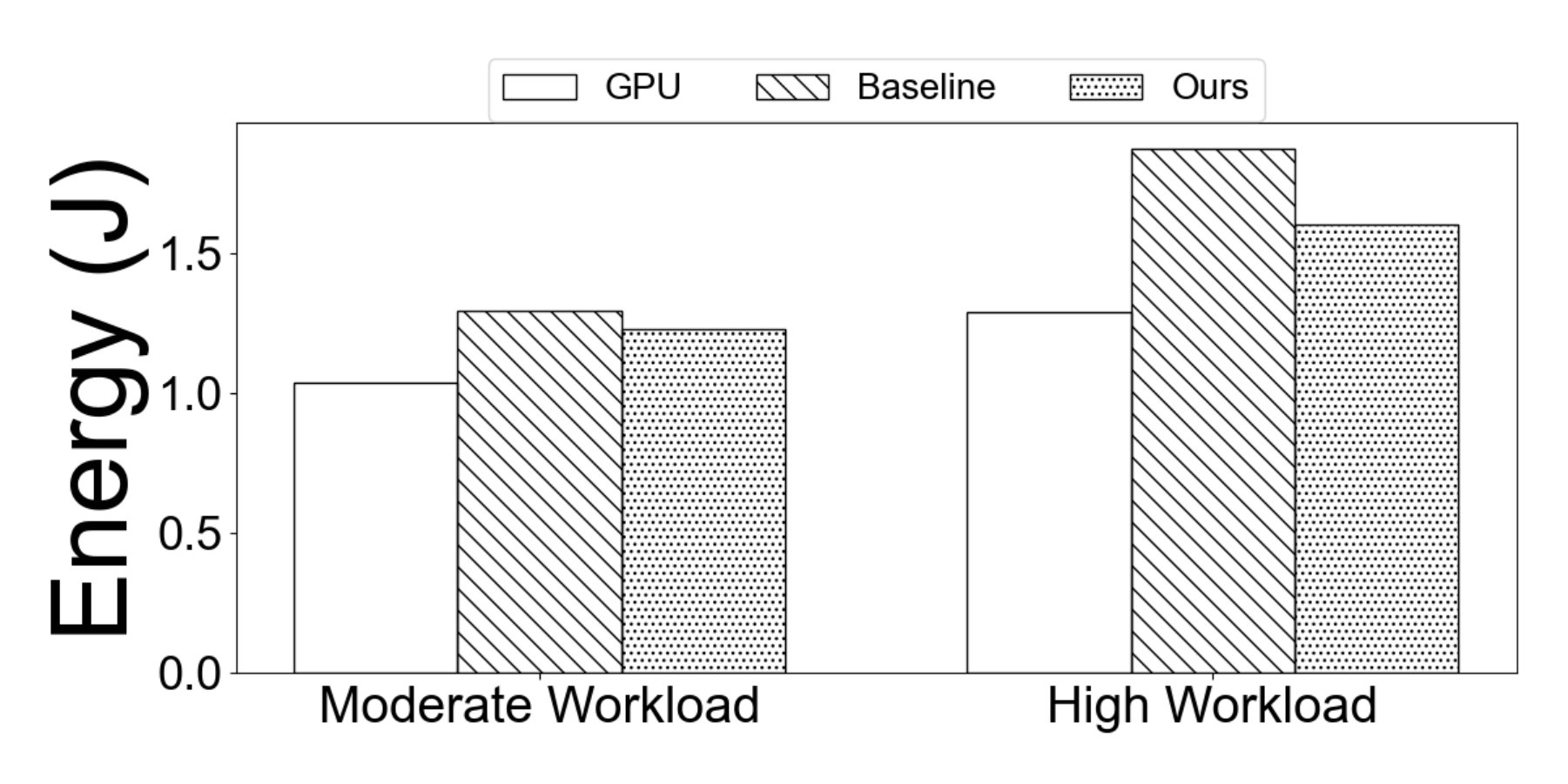}
    \label{fig:Energy}
  }
  \vspace{-3mm}
  % \vspace{-1mm} % Adjust vertical space if needed
  \caption{Performance comparison under different workload conditions.}
  % \vspace{-3mm} % Adjust vertical space if needed
  \label{fig:async-sync compare}
\end{figure}
\vspace{-5mm}
\lsc{We compare the performance of MACE on GPU, CoDL, and \sysname with YOLO v2 on Xiaomi9 with Snapdragon855 SoC with two workload conditions, moderate and high.
For the moderate workload condition, we configure the CPU frequency to 1.49 GHz, GPU frequency to 499 MHz, and average CPU utilization to 78.8\%. 
Conversely, for the high workload condition, the CPU frequency is set to 0.88 GHz, and the GPU frequency to 427 MHz, with an average CPU utilization of 91.3\%.
Figure \ref{fig:async-sync compare} illustrates the results of our experiments. Compared to CoDL~\cite{CoDL2022}, \sysname demonstrates a 3.94\% and 12.97\% reduction in latency, along with an enhancement in energy efficiency by 4.06\% and 16.88\% for moderate and high workload conditions, respectively.
}
\section{CONCLUSION}
\label{sec:intro}

\lsc{
This paper introduces \sysname, a responsive and energy-efficient concurrent deep neural network (DNN) inference system designed for mobile devices. 
Comprising a runtime energy profiler and an energy-aware operator partition module, \sysname addresses the challenge of accurately predicting hardware-related energy cost in dynamic conditions and fast adaptation of computation partition.
% by intelligently determining workload distribution across heterogeneous processors.
Experimental evaluations conducted on real-world mobile devices, with various execution schemes and workloads, demonstrate the effectiveness of \sysname. 
Results indicate a reduction in energy cost, up to 16.88\%, and a decrease in latency by 12.97\% compared to CoDL, particularly under high workload conditions.
}

\section*{Acknowledgements}
This work was partially supported by the National Science Fund for Distinguished Young Scholars (62025205) and the National Natural Science Foundation of China (No. 62032020, 62102317).
%The authors thank the anonymous reviewers for their constructive feedback that has made the work stronger.

%\newpage
\bibliography{acmart}
\bibliographystyle{ACM-Reference-Format}
\end{document}